# SUPPRESSION OF PLASMA INSTABILITIES IN SPACE AND GROUND NUCLEAR POWER DEVICES

**A.S. Mustafaev, A.Yu. Grabovskiy, B.D. Klimenkov**

The conditions for the excitation of current and voltage oscillations in the plasma of a three-electrode current and voltage stabilizer are experimentally investigated. It was found that in the regimes under consideration, the plasma has negative conductivity, which causes oscillations. A highly efficient method for suppressing plasma instabilities is proposed, based on controlling the sign of differential plasma conductivity by adjusting the concentration of slow electrons using an external electrode. The proposed method makes it possible to achieve a high level of stability of the energy characteristics of the stabilizer at a discharge current density of up to 5 A/cm2 and a power of 500 W/cm2.

# ПОДАВЛЕНИЕ ПЛАЗМЕННЫХ НЕУСТОЙЧИВОСТЕЙ В ПРИБОРАХ КОСМИЧЕСКОЙ И НАЗЕМНОЙ ЯДЕРНОЙ ЭНЕРГЕТИКИ

© А.С.-У. Мустафаев, А.Ю. Грабовский, Б.Д. Клименков

[1]*Санкт-Петербургский горный университет,*
*E-mail: schwer@list.ru*

Экспериментально исследованы условия возбуждения колебаний тока и напряжения в плазме трехэлектродного стабилизатора тока и напряжения. Установлено, что в рассматриваемых режимах плазма обладает отрицательной проводимостью, которая и вызывает колебания. Предложен высокоэффективный метод подавления плазменных неустойчивостей, основанный на управлении знаком дифференциальной проводимости плазмы путем регулировки концентрации медленных электронов с помощью внешнего электрода. Предлагаемый метод позволяет достигать высокого уровня стабильности энергетических характеристик стабилизатора при плотности разрядного тока до 5 А/см$^2$ и мощности 500 Вт/см$^2$.
Ключевые слова: зондовый метод, функция распределения электронов, электрокинетические характеристики, трехэлектродный стабилизатор напряжения, плазменные колебания, отрицательная проводимость.

## ВВЕДЕНИЕ

Одним из ключевых направлений развития систем вооружений в Российской Федерации является разработка малогабаритных сверхмощных ядерных энергетических установок (ЯЭУ) [1]. Для решения этой задачи требуются электронные компоненты, позволяющие реализовать полное управление током в цепях ЯЭУ. К таким компонентам (приборам) выставляются следующие требования на допустимые уровни излучений реактора на стенке приборного отсека: флюенс быстрых нейтронов ($E_n > 0{,}1$ МэВ) должен быть $\leq 10^{12}$ н/см$^2$; поглощенная доза фотонов (γ-квантов) должна быть $\leq 10^6$ рад.

Таким требованиям удовлетворяют твердотельные полупроводниковые приборы, но с ростом электрической мощности и рабочей температуры применение для них локальной радиационной защиты резко ухудшает массогабаритные характеристики ЯЭУ. В этих условиях единственным решением является применение газоразрядных электронных приборов на базе неравновесной плазмы [2, 3].

Эксплуатация приборов плазменной энергетики сопряжена с проблемой возбуждения неустойчивостей при попытках повышения их энергетических параметров [4, 5]. Колебания и неустойчивости такого типа можно использовать в практических целях, однако их генерация в условиях плазменных источников и стабилизаторов оказывает разрушительное воздействие на их энергетические и электрокинетические характеристики. Очевидно, что борьба с неустойчивостями подразумевает управление функцией распределения электронов по скоростям (ФРЭС) в рабочих режимах плазменных приборов.

В настоящей работе для решения проблемы подавления неустойчивостей в плазменных приборах разработана специальная конструкция триода на базе гелиевого низковольтного пучкового разряда (НПР) с управляющим электродом (УЭ), расположенным вне зазора катод-анод. Такая конструкция позволяет управлять функцией распределения заряженных частиц для эффективного подавления возникающих плазменных неустойчивостей и колебаний.

## ЭКСПЕРИМЕНТАЛЬНЫЕ УСТАНОВКА, ПРИБОР И МЕТОД ИССЛЕДОВАНИЯ

Схема экспериментальной установки и конструкция плазменного триода представлены на рис. 1 и 2 соответственно. Их детальное описание можно найти в работе [2]. Рассмотрим основные элементы установки. В блок I входят вакуумная камера (1), катодный (2) и анодный (3) узлы. Специальные окна из сапфира (4) использовались для наблюдений и оптических измерений. Вакуумная система (блок II) включает турбомолекулярный насос, обеспечивающий максимально достижимое разрежение 5·10$^{-8}$ тор. Высокая стабильность разрядных условий достигалась длительным циклом обезгаживания при температуре 700 К.

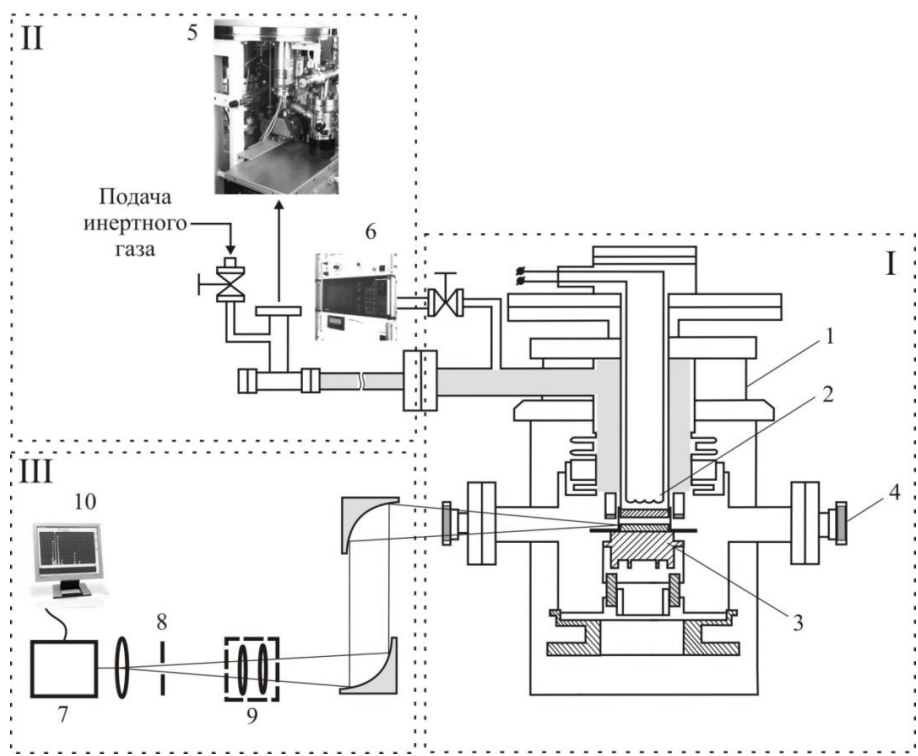

**Рис. 1.** Принципиальная схема экспериментальной установки; 1 - вакуумная камера, 2 - катодный узел, 3 -анодный узел, 4 - сапфировое окно, 5 -турбомолекулярный насос, 6 - комплекс масс - спектрометрического анализа, 7 - монохроматор, 8 - диафрагма, 9 - конденсор, 10 -система обработки экспериментальных данных [2].

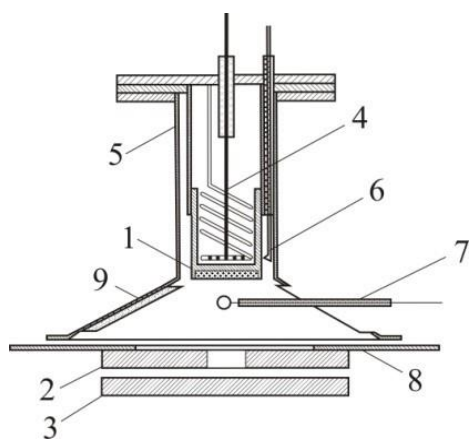

**Рис. 2.** Конструкция плазменного триода: 1 - катод; 2 - анод; 3 - УЭ; 4 - нагреватель; 5 - тепловой экран; 6 - катодная микротермопара; 7 - плоский зонд; 8 – защитные изоляторы; 9 - боковой проводящий экран [2].

Схема экспериментального прибора представлена на рис. 2. Его катод (1) изготовлен из пористого вольфрама, импрегнированного барием и нагревается тантал-ниобиевой проволокой. Анод (2) и УЭ (3) выполнены из поликристаллического молибдена. Диаметр катода 10 мм, диаметры анода и УЭ равны и составляют 30 мм. Компоновка триода нетрадиционна – анод имеет центральное отверстие диаметром 2 мм, а УЭ вынесен из межэлектродного зазора и располагается непосредственно за анодом. Зазор катод-анод составляет 8 мм, УЭ расположен на расстоянии 1 мм за анодом.

Плазменный канал ограничен металлическим конусным экраном (9), находящимся под потенциалом катода.

Диагностика плазмы НПР осуществлялась методом плоского одностороннего зонда, развитым в работах [6-8]. Поскольку в рассматриваемых экспериментальных условиях плазма обладает осевой симметрией, ФРЭС не зависит от азимутального угла (рис. 3):

$$f(\vec{r}, \vec{\upsilon}) = f(\vec{r}, \upsilon, \theta), \qquad (1)$$

где $\upsilon = |\vec{\upsilon}|$; $\theta$ - полярный угол. Ток на плоский ленгмюровский зонд:

$$I = qS \int \upsilon_n f(\vec{\upsilon}) d\vec{\upsilon} = \frac{2qS}{m^2} \int_0^{2\pi} d\phi' \int_{qU}^{\infty} \varepsilon d\varepsilon \int_0^{\theta'_{\max}} f(\varepsilon, \theta', \phi') \cos\theta' \sin\theta' d\theta', \qquad (2)$$

где $\upsilon_n$ - нормальная к поверхности зонда составляющая вектора скорости электрона $\vec{\upsilon}$, $U$ - потенциал зонда; $\varepsilon = m\upsilon^2/2$; $\phi'$ и $\theta'$ - соответственно азимутальный и полярный углы вектора $\vec{\upsilon}$. Вторая производная зондового тока (2) по потенциалу зонда $U$:

$$I''_U = \frac{q^3 S}{m^2} \left[ \int_0^{2\pi} f(qU, \theta' = 0, \phi') d\phi' - \int_0^{2\pi} d\phi' \int_{qU}^{\infty} \frac{\partial}{\partial (qU)} f(\varepsilon, \theta'_{\max}, \phi') d\varepsilon \right]. \qquad (3)$$

Выражение (3) можно переписать в виде

$$I''_U(qU, \alpha) = \frac{2\pi q^3 S}{m^2} \left[ f(qU, \alpha) - \frac{1}{2\pi} \int_0^{2\pi} d\phi' \int_{qU}^{\infty} \frac{\partial}{\partial (qU)} f(\varepsilon, \theta^*) d\varepsilon \right]. \qquad (4)$$

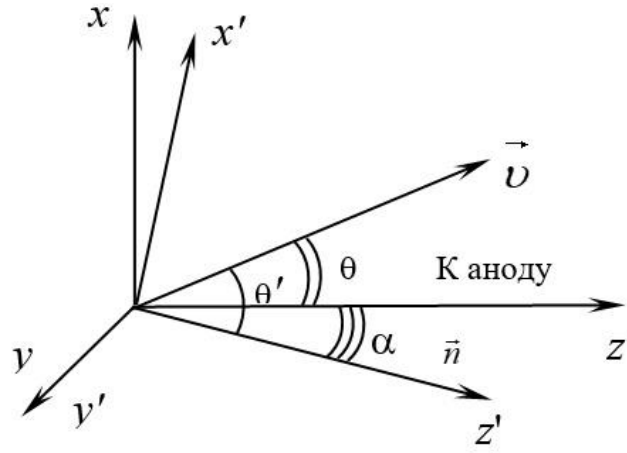

**Рис. 3.** Ориентация зонда в плазме.

Для нахождения ФРЭС, представим $f(\varepsilon, \theta)$ и $I''_U(qU, \alpha)$ в виде ряда по полиномам Лежандра:

$$f(\varepsilon, \theta) = \sum_{j=0}^{\infty} f_j(\varepsilon) P_j(\cos\theta) \tag{5}$$

$$I''_U(qU, \alpha) = \frac{2\pi q^3 S}{m^2} \sum_{j=0}^{\infty} F_j(qU) P_j(\cos\alpha). \tag{6}$$

После подстановки (5) и (6) в (4) получаем соотношение между компонентами $f_j$ и $F_j$:

$$f_j(qU) = F_j(qU) + \int_{qU}^{\infty} f_j(\varepsilon) \frac{\partial}{\partial(qU)} P_j\left(\sqrt{\frac{qU}{\varepsilon}}\right) d\varepsilon. \tag{7}$$

Выражение (7) является интегральным уравнением Вольтерра II рода. Решая его с помощью метода резольвент [9], получим

$$f_j(qU) = F_j(qU) + \int_{qU}^{\infty} F_j(\varepsilon) R_j(qU, \varepsilon) d\varepsilon, \tag{8}$$

Подставив в уравнение (8) соотношение

$$F_j(qU) = \frac{(2j+1)m^2}{4\pi q^3 S} \int_{-1}^{1} I''_U(qU, x) P_j(x) dx$$

и получим основную формулу, позволяющую реконструировать компоненты $f_j$:

$$f_j(qU) = \frac{(2j+1)m^2}{4\pi q^3 S} \int_{-1}^{1} \left[ I''_U(qU, x) + \int_{qU}^{\infty} I''_U(\varepsilon, x) R_j(qU, \varepsilon) d\varepsilon \right] P_j(x) dx. \tag{9}$$

Таким образом, для диагностики плазмы методом плоского зонда необходимо зарегистрировать значения $I_U''(qU, \alpha)$ при различных углах зонда относительно оси разряда, рассчитать по формуле (9) ряд лежандровых компонентов и реконструировать полную ФРЭС согласно формуле (5). Формула (9) демонстрирует, что метод не требует априорной информации об анизотропии распределения заряженных частиц, однако для корректного описания ФРЭС в сильнонеравновесной плазме требуется большое число компонент, что не всегда возможно в реальном эксперименте. В этой связи метод плоского зонда усовершенствован авторами для диагностики моноэнергетичных пучков заряженных частиц [10].

В процессе зондовых измерений тщательно выдерживались все требования, теории Ленгмюра. С этой целью изготавливались зонды толщиной 30 мкм и диаметром 0,5 мм. Также учитывались все факторы, способные искажать зондовые ВАХ [11].

Для получения значений второй производной зондового тока использовался метод демодуляции, реализованный в измерительном комплексе на базе ПК. В качестве дифференцирующего сигнала, для увеличения чувствительности метода использовалось 100 % модулированное напряжение $\Delta U = U_0(1 + \cos\omega_1 t)\cos\omega_2 t$ [11].

## ЭКСПЕРИМЕНТАЛЬНЫЕ РЕЗУЛЬТАТЫ И ИХ ОБСУЖДЕНИЕ

**Распределение параметров в плазме НПР**

В эксперименте роль плазмообразующего компонента играл гелий, поскольку он обладает самыми высокими среди других инертных газов потенциалами ионизации и возбуждения. Это позволяет добиться наиболее яркого проявления нелокальных эффектов в плазме. Экспериментальные исследования проводились в диапазоне токов 0,1 - 2 А и давлений гелия 0,1 - 10 тор. Выбор таких условий обусловлен возможностью перевода плазмы как в локальный, так и в нелокальный режим. В последнем случае длины релаксации на электрон-электронных $L_{ee}$ и электрон-атомных $L_{ea}$ столкновениях превосходят длину межэлектродного зазора $d$ [12, 13].

Вторые производные $I_U''$ плоского зонда, зарегистрированные при его различных ориентациях относительно оси разряда при $P_{He}$=2,5 тор показаны на рис. 4а. Видно, что структура ФРЭС формируется двумя обособленными группами электронов: тепловыми $f_t$ и пучковыми $f_0$ с концентрацией и средней энергией $n_t$, $\bar{\varepsilon}_t$ и $n_0$, $\bar{\varepsilon}_0$ соответственно.

Распределение тепловых электронов по скоростям близко к максвелловскому, тогда как группа быстрых электронов отличается значительной анизотропией распределения. Сравнение энергий тепловых и пучковых электронов выявляет сильную неравновесность ФРЭС по энергии - энергия тепловых электронов составляет приблизительно 2 эВ, быстрых – приблизительно 25 эВ.

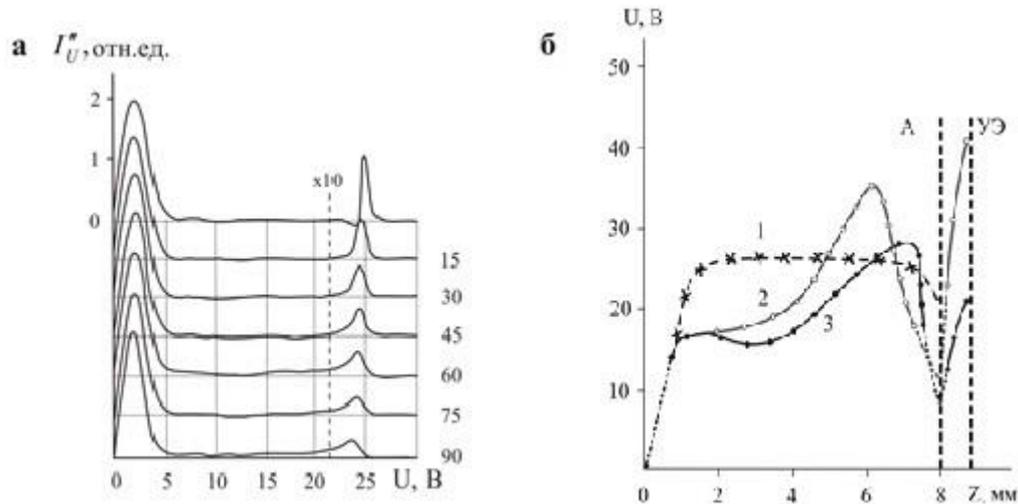

**Рис. 4.** Вторые производные $I_U''$ ($qU$, $\alpha$), полученные при различных углах плоского зонда относительно оси разряда (а); $P_{He}$ = 2,5 тор, $j$ = 0,15 А/см², $U$ = 24 В, $n_t$ = 1,97·10¹¹ см⁻³, $n_0$ = 3,76·10⁹ см⁻³ и распределение потенциала по зазору экспериментального прибора (б); $P_{He}$, тор: 1 - 2; 2 - 3; 3 - 6; ток на анод $i$, А: 1 - 0,2; 2 - 0,6; 3 - 0,4; ток на УЭ $i_{уэ}$, А: 1 - 0; 2 – 0,04; 3 – 0,02.

Формирование такой структуры ФРЭС связано с распределением потенциала по межэлектродному зазору (рис. 4б, кривая 1). Видно, что у анода образуется скачок потенциала $\varphi_a \approx 1{,}5\bar{\varepsilon}_t$, являющийся практически непреодолимым потенциальным барьером для тепловых электронов плазмы. Электроны, эмитированные катодом, ускоряются на прикатодном скачке потенциала и образуют пучок с небольшим разбросом по энергии, пронизывающий межэлектродный зазор. Рождение медленных электронов происходит в результате ионизации атомов гелия электронами пучка. Потенциалы возбуждения и ионизации для $He$ составляют $U_м \approx 19{,}8$ В и $U_{ион} \approx 24{,}6$ В соответственно, что обуславливает доминирующую роль пучка в неупругих процессах и токопереносе.

Если рассматривать промежуток между анодом и управляющим электродом, то отверстие в аноде является некоторым подобием плазменного катода, и токоперенос здесь осуществляется в основном тепловыми электронами из плазмы в зазоре катод-анод, что становится возможным благодаря их ускорению в сильном электрическом поле двойного слоя в окрестностях анода (рис. 4б, кривые 2, 3).

С целью изучения процессов, протекающих в рассматриваемых плазменных условиях по экспериментально зарегистрированным значениям $I_U''(qU,\alpha)$ восстановлены Лежандровы компоненты разложения ФРЭС $f_0 - f_6$, определяющие набор важнейших базовых параметров плазмы: концентрацию, плотность электронного тока, анизотропию электронного давления, и др. [11]. В частности, по компоненту $f_0$ можно рассчитать функцию возбуждения атомов и генерации ионов:

$$\Gamma = 4\pi N_a \int_{v_{nop}}^{\infty} v\sigma_{ea}^i(v)f_0(v)v^2 dv, \qquad (10)$$

где $\sigma_{ea}^i$ - энергетический ход сечения соответствующего процесса, $\upsilon$ - скорость налетающего электрона.

**Подавление колебаний тока и напряжения в плазменных приборах с отрицательной проводимостью**

Известно, что НПР может быть неустойчив к возбуждению различного типа колебаний [12-14], разрушающих рабочие режимы плазменных электронных приборов.

ВАХ гелиевого НПР, зарегистрированные в условиях трехэлектродного стабилизатора тока и напряжения представлены на рис. 5. Видно, что ВАХ 1 имеет участок отрицательной дифференциальной проводимости $G_д$, что и является причиной развития колебаний разрядного тока и напряжения [15].

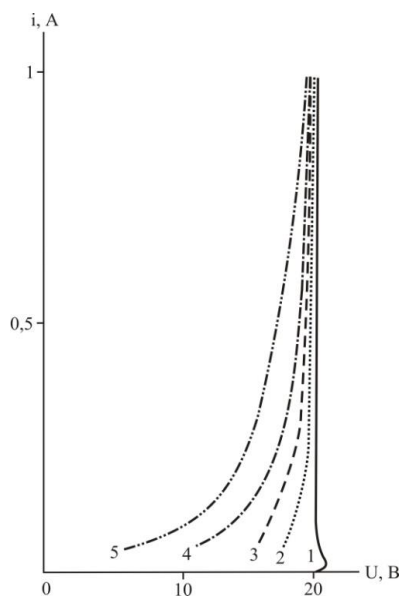

**Рис. 5.** ВАХ стабилизатора при различных величинах тока на УЭ; $i_{уэ}$, А: 1 - 0; 2 - 0,02; 3 - 0,04; 4 - 0,06; 5 - 0,08. $P_{He} = 3$ тор.

Типичная форма таких колебаний представлена на рис. 6. Видно, что колебания происходят при полной модуляции тока в разряде на частотах 50-150 кГц и при амплитудных значениях напряжения порядка 30 В. В условиях постоянного давления частота колебаний практически не изменяется при варьировании уровня разрядного тока. Если уровень тока возрастает, то глубина модуляции падает вплоть до 10 %. При увеличении давления наполнителя частота колебаний нарастает линейно (рис. 7).

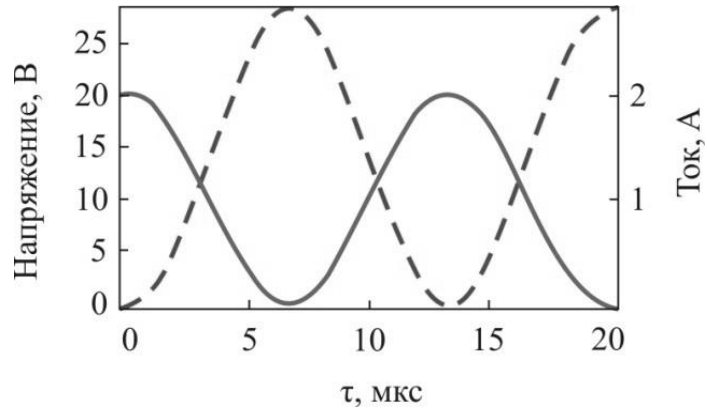

**Рис. 6.** Форма колебаний тока разряда (сплошные линии) и напряжения (пунктир) при $P_{He}$ = 5 тор.

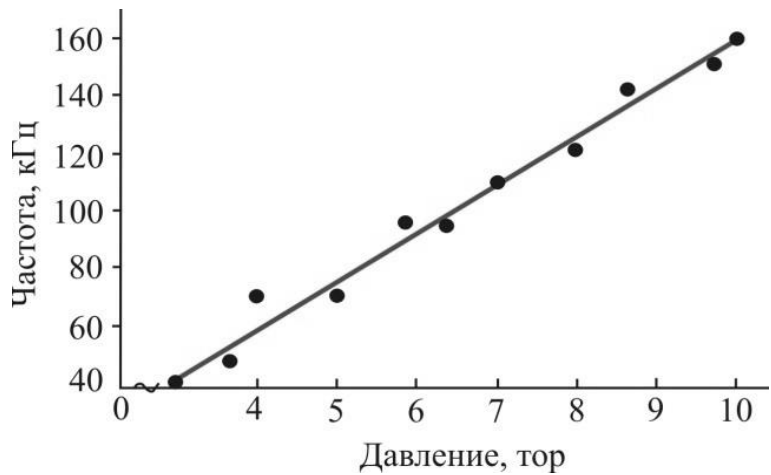

**Рис. 7.** Частота колебаний в зависимости от тока от давления гелия.

Аксиальное распределение параметров плазмы перед возникновением колебаний, приведено в таблице 1. Функция генерации $\Gamma$ (показывает число ионов, родившихся в единице объёма, за 1 сек) рассчитывалась по формуле (10) с помощью измеренного лежандрова компонента ФРЭС $f_0$. Величина $D$ определяет уход ионов из плазменного канала в радиальном направлении и принимается равной

$$D = n/\tau,$$

где $\tau$ - время радиальной диффузии ионов ($\tau \sim r/\overline{\upsilon}_{др}$). Скорость движения ионов в радиальном поле напряженностью $E_r \sim 5$ В/см составляла величину порядка $\overline{\upsilon}_{др} = 2 \cdot 10^5$ см/сек [16].

Таблица 1. Аксиальное распределение параметров плазмы. $P_{He}$=2 тор, разрядный ток $i$=0,5 А

| $Z$, мм | $\varphi$, В | $n_0$, $10^9$ см$^{-3}$ | $n_t$, $10^{11}$ см$^{-3}$ | $\tau_{диф}$, с | $D \sim n/\tau$, $10^{17}$ см$^{-3}$·с$^{-1}$ | $\Gamma$, $10^{17}$ см$^{-3}$·с$^{-1}$ | $\Gamma/D$ |
|---|---|---|---|---|---|---|---|
| 1 | 20,1 | 15 | 0,5 | 4,0 | 0,1 | 6,0 | 60,0 |
| 2 | 20,7 | 12 | 0,75 | 5,5 | 0,13 | 4,8 | 37,0 |
| 3 | 21,2 | 7 | 1,23 | 7,0 | 0,17 | 2,8 | 16,4 |
| 4 | 21,4 | 4 | 1,6 | 8,5 | 0,2 | 1,6 | 8,0 |
| 5 | 22 | 2,74 | 2,5 | 10 | 0,24 | 1,1 | 4,58 |
| 6 | 22,4 | 1,75 | 3,73 | 11,5 | 0,27 | 0,7 | 2,59 |
| 7 | 22,8 | 1,11 | 5,57 | 13 | 0,3 | 0,44 | 1,47 |
| 8 | 23,3 | 0,71 | 8,29 | 14,5 | 0,34 | 0,28 | 0,82 |

Из таблицы 1 видно, что по мере удаления от катода в аксиальной области плазмы несколько возрастает член $D$ по причине увеличения радиального градиента концентрации и поля. Вместе с тем, вследствие влияния радиальной диффузии и неупругих процессов происходит обеднение быстрой части ФРЭС, влекущее снижение функции генерации ионов $\Gamma$. Вблизи анода величины $\Gamma$ и $D$ сопоставимы (рис. 8), что дает возможность записать условие стационарности концентрации ионов в виде $\Gamma \approx D$. В случае нарушения условия, когда $\Gamma \geq D$, формируется избыточное количество ионов в прианодном регионе с последующим возрастанием тока медленных электронов на анод и возникают релаксационные колебания, характер которых зависит еще и от анодной нагрузки. При $\Gamma \leq D$ дефицит ионов может приводить к обрыву тока.

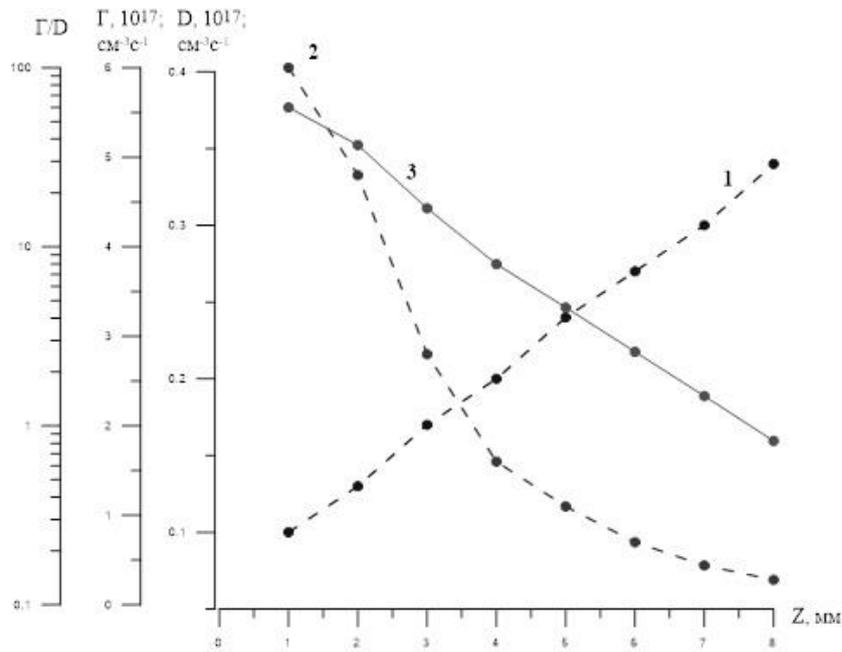

**Рис. 8.** Аксиальные зависимости величин *D* (1), *Г* (2) и *Г/D* (3).

Таким образом, физическую картину возбуждения колебаний можно сформировать, рассматривая соотношение процессов генерации ионов в плазме и их ухода из межэлектродного зазора. С ростом давления область максимальной концентрации ионов смещается к аноду, что ведет к увеличению диффузионного потока ионов на анод. В результате происходит компенсация частично отрицательного объемного заряда в прианодной области. Следствием такого взаимодействия является рост хаотического тока медленных электронов, ранее запертых в потенциальной яме между катодом и анодом. Амплитудное значение тока задается хаотическим током тепловых электронов. В предположении нулевого прианодного потенциального барьера его величину можно записать в виде:

$$i_t = \frac{1}{4} e n_e \upsilon_t S,$$

где $e$ - заряд электрона, $n_e$ - концентрация электронов вблизи анода, $\upsilon_t$ - тепловая скорость электронов, а $S$ - площадь анода. Оценки при $P_{He}$=5 тор, $n_e \sim 10^{11}$ см$^{-3}$, $T_e \sim 1$ эВ дают $i_t \sim 1,8$ А, что приблизительно соответствует зарегистрированному значению (рис. 6).

На основании проведенных исследований предложен способ подавления колебаний тока и напряжения [17]. Из рис. 5 видно, что отбор тока тепловых электронов на управляющий электрод позволяет управлять знаком $G_д$. Возможность такого управления позволяет эффективно подавлять колебания тока и напряжения, поскольку во всех режимах с положительной дифференциальной проводимостью (рис. 5, кривые 2-5) колебания отсутствовали и более не возбуждались.

ЗАКЛЮЧЕНИЕ

В работе предложена конструкция прибора и способ, заключающийся в управлении знаком дифференциальной проводимости для эффективной борьбы с плазменными колебаниями и неустойчивостями. Способ является универсальным, и при его практической реализации были созданы эффективные радиационностойкие газоразрядные приборы для космической и наземной ядерной энергетики: стабилизаторы, инверторы, ключевые элементы, генераторы, и др., в которых проблема плазменных неустойчивостей была решена полностью и во всем диапазоне рабочих параметров. В условиях высокой стабильности плазменных условий удалось создать высокочувствительные портативные микроплазменные газоанализаторы, предназначенные для решения широкого круга задач: от анализа состава атмосферы на предприятиях нефтегазовой промышленности и горной отрасли до ранней диагностики заболеваний по летучим биомаркерным молекулам в дыхании человека [18].